\documentclass[lettersize,journal]{IEEEtran}
\usepackage{amsmath,amsfonts}
\usepackage{algorithmic}
\usepackage{multirow}
\usepackage{algorithm}
\usepackage{array}
\usepackage[caption=false,font=normalsize,labelfont=sf,textfont=sf]{subfig}
\usepackage{textcomp}
\usepackage{stfloats}
\usepackage{url}
\usepackage{verbatim}
\usepackage{graphicx}
\usepackage{cite}
\usepackage{xcolor}
\begin{document}

\title{Iterative Approach to Reconstructing Neural Disparity Fields from Light-field Data}

\author{Ligen Shi, Chang Liu, Xing Zhao*, and Jun Qiu*
\thanks{Ligen Shi and Xing Zhao are with the School of Mathematical Sciences, Capital Normal University, Beijing 100048, China, (ligenshi0826@gmail.com, zhaoxing\_1999@126.com)}
\thanks{Chang Liu and Jun Qiu are with College of Computer Science, Beijing Information Science and Technology University, Beijing 100101, China, (changliuct@gmail.com, qiujun@bistu.edu.cn)}
\thanks{Corresponding author: Xing Zhao and Jun Qiu.}}

\markboth{Journal of \LaTeX\ Class Files,~Vol.~x, No.~x, x~2024}%
{Shell \MakeLowercase{\textit{et al.}}: A Sample Article Using IEEEtran.cls for IEEE Journals}

\IEEEpubid{0000--0000/00\$00.00~\copyright~2021 IEEE}

\maketitle

\begin{abstract}
This study proposes a neural disparity field (NDF)  that establishes an implicit, continuous representation of scene disparity based on a neural field and an iterative approach to address the inverse problem of NDF reconstruction from light-field (LF) data. NDF enables seamless and precise characterization of disparity variations in three-dimensional scenes and can discretize disparity at any arbitrary resolution, overcoming the limitations of traditional disparity maps that are prone to sampling errors and interpolation inaccuracies. The proposed NDF network architecture utilizes hash encoding combined with multilayer perceptrons (MLPs) to capture detailed disparities in texture levels, thereby enhancing its ability to represent the geometric information of complex scenes. By leveraging the spatial-angular consistency inherent in the LF data, a differentiable forward model to generate a central view image from the LF data is developed. Based on the forward model, an optimization scheme for the inverse problem of NDF reconstruction using differentiable propagation operators is established. Furthermore, an iterative solution method is adopted to reconstruct the NDF in the optimization scheme, which does not require training datasets and applies to LF data captured by various acquisition methods. Experimental results demonstrate that the proposed method can reconstruct high-quality NDF from LF data. The high-resolution disparity can be effectively recovered by NDF, demonstrating its capability for the implicit, continuous representation of scene disparities.
\end{abstract}


\section{Introduction}
\IEEEPARstart{C}{omputational} imaging incorporates computer vision and signal processing techniques into imaging systems to reconstruct high-dimensional optical signals that carry scene information through novel imaging mechanisms, optical paths, and reconstruction methods, thereby achieving efficient and high-precision observation of scenes \cite{2006LevoyLF}. Computational imaging surpasses the limitations of classical imaging in terms of signal dimensions, scale, and resolution.

From the perspective of imaging methodology, light-field (LF) imaging operates as a close-proximity passive direct line-of-sight imaging technology that relies on ambient light to capture scene content within its visual ambit. The characteristic features include narrower baselines, increased uniformity in viewpoint sampling, and diminished occlusions. Notably, substantial domain discrepancies emerge in the LF data generated by different cameras owing to inherent variations in their imaging modalities, structural designs, and internal parameters.

Deep learning-based reconstruction methods rely on labeled training data and do not explore the inherent consistency within LF data. As a result, networks trained with data from one camera model are not generalizable to other camera models. Since individual LF data contain abundant information, exploiting their spatial-angular consistency allows a network to reduce its reliance on specific camera models, thereby promoting generalization across different camera models.

The neural field \cite{2020NeRF, NeRFPPF, NeFieldsinVisualComputingBeyond} represents the scene to be reconstructed as a continuous implicit function, achieving high-quality visual results through neural network parameterization. The continuous implicit function representation ensures continuity and is not limited by resolution constraints. By employing neural network parameterization of signal values, this representation method adopts a `discrete-to-continuous' computational approach and utilizes differentiability for solving inverse problems. Compared to traditional methods like explicit pixel- or matrix-based representations, neural fields offer higher precision, lower computational complexity, and greater robustness. They achieve this by modeling scenes continuously, capturing details without resolution constraints. In contrast, traditional methods often suffer from aliasing or detail loss at higher resolutions. Neural fields are also more storage-efficient, utilizing implicit functions that reduce memory demands for high-resolution data. By focusing on the forward problem and using automatic differentiation to learn the inverse, neural fields adapt better to varying sampling patterns. Unlike deep learning methods, which require large labeled datasets, neural fields optimize individual scenes without such data. Their models are based on physical models rather than relying on `black-box' mappings, allowing for improved generalization across datasets and better performance with new or unseen data.
\IEEEpubidadjcol

\begin{figure}[!t]
\centering
\includegraphics[width=3.2in]{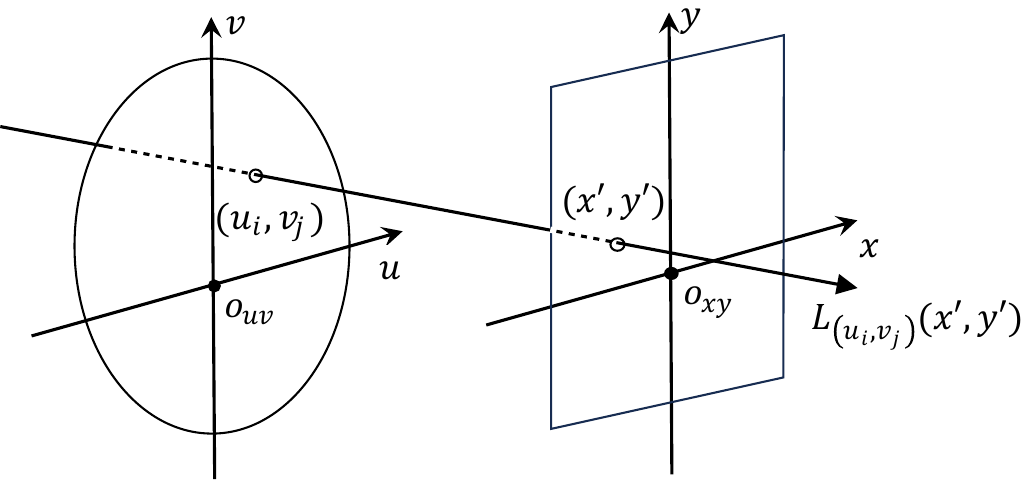}
\caption{The four-dimensional two-plane LF model.}
\label{Fig_1}
\end{figure}

The commonly used four-dimensional (4D) two-plane LF model\cite{levoy2023light, ng2005light,xiao2013advances} is shown in Fig.~\ref{Fig_1}, where $(u, v)$ represents the viewpoint plane and $(x, y)$ represents the image plane. A pixel $(x', y')$ at viewpoint $(u_i, v_j)$ is expressed as $L_{u_i, v_j}(x', y')$. A key feature of LF data is that all viewpoints lie on the same horizontal plane, with very small baselines between them. This small baseline makes three-dimensional (3D) reconstruction challenging, especially when using volume rendering methods\cite{max1995optical,2020NeRF}.

\begin{figure*}[!t]
\centering
\includegraphics[width=15cm]{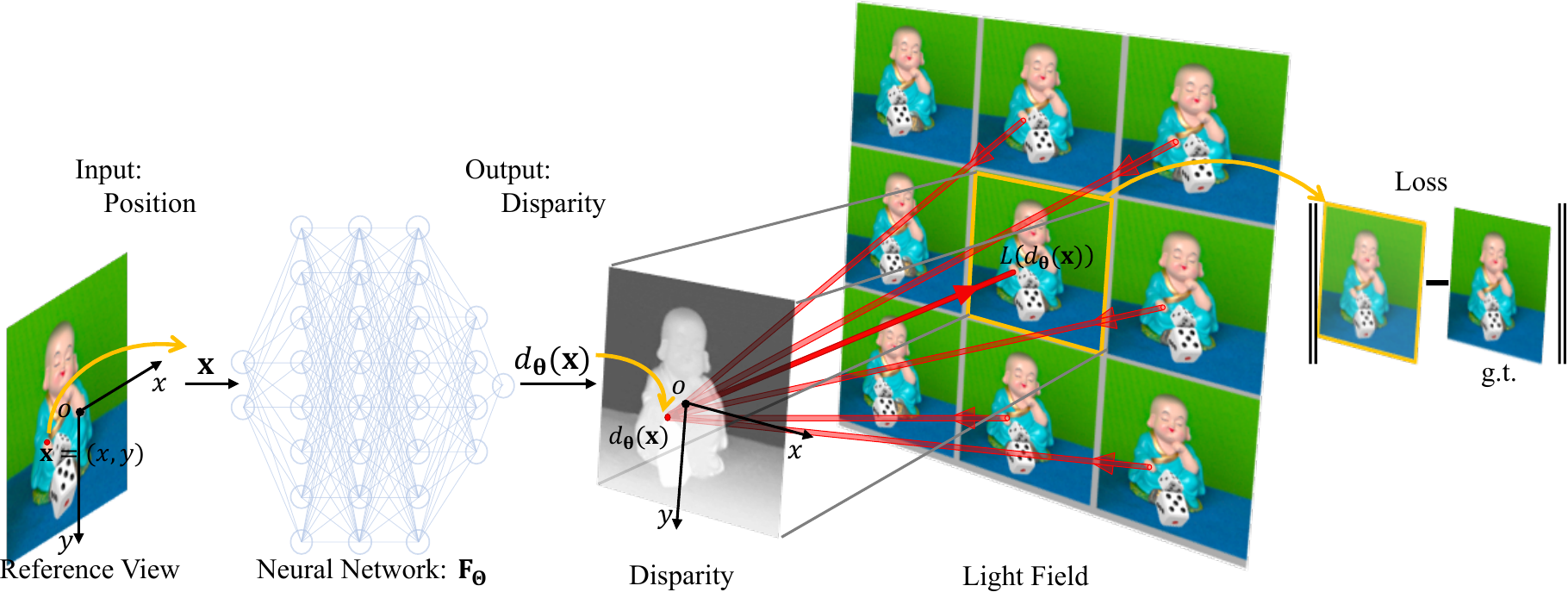}
\caption{Framework for the implicit representation of neural disparity fields. Red ray bundles indicate the propagation of disparities from the visible views towards the central view.}
\label{Fig_2}
\end{figure*}

Therefore, we leverage the spatial-angular consistency within individual LF data to implicitly represent the disparity contained within it and construct a neural disparity field (NDF). The inverse problem-solving model based on the NDF was established to reconstruct the disparity from the LF. This model was trained and iteratively solved through differentiability, effectively transforming the disparity estimation problem into an optimization problem for the NDF. The framework for the implicit representation of an NDF is shown in Fig.\ref{Fig_2}.

Unlike deep learning-based methods for disparity estimation, the NDF reconstruction method does not depend on training datasets. It can be applied to two-plane parameterized LF data using any collection method. By utilizing the implicit representation of the neural field, this disparity reconstruction method departs from the traditional approach of representing disparity as discrete points, overcoming the inevitable sampling, interpolation, and other discretization calculation issues of explicit discrete disparity and establishing a continuous disparity surface with implicit expression. The innovations of this study are as follows:
\begin{enumerate}
\item {Disparity is parameterized using neural networks to establish an NDF, that creates an implicit continuous representation of scene disparity functions. This NDF enhances the expressive capability of complex details of the disparity surface and effectively reduces computational errors introduced by discretization. Unlike traditional hierarchical disparity search estimation methods, a representation based on the NDF allows the formation of a differentiable propagation forward problem model, inverse problem optimization modeling, and iterative solving of the reconstruction algorithm for continuous implicit disparity functions.}
\item {From the perspective of spatial-angular consistency, a forward problem model for the disparity propagation of the NDF under non-ideal scene conditions (with occlusion and noise) is constructed. This forward-problem model, described by differentiable propagation operators, can propagate any viewpoint image to a central view via the NDF.}
\item{Based on the forward problem model, an optimization model for the inverse problem of the NDF was established using differentiable propagation operators. An iterative approach was proposed to solve the disparity reconstruction from the LF. This method offers a novel solution to the disparity reconstruction problem. The iterative reconstruction method based on the inverse problem does not rely on training datasets and can be applied to two-plane parameterized LF data obtained by any collection method. Furthermore, it allows the application of inverse problem-solving strategies in disparity reconstruction.}
\end{enumerate}

\section{Related work}
We review optimization-based and deep learning-based methods for LF depth estimation, as well as recent developments in neural fields.
\subsection{Optimization-based traditional methods}
Disparity estimation algorithms based on traditional optimization methods adhere to a computational approach known as `discrete-to-discrete'. These algorithms aim to estimate corresponding discrete disparity images from discretized LF data. However, this approach is influenced by resolution limitations and inevitably introduces various errors.

Early disparity estimation methods explore the structural information of LF data to assess consistency among different views. Wanner et al. \cite{wanner2013variational} introduce structural tensors for depth estimation. Heber et al. \cite{2014PCA} employ principal component analysis for stereo matching, focusing on subview images that impact matching errors while neglecting those with minor influences. To overcome the challenges of matching small baselines between subview images in LFs, Jeon et al. \cite{jeon2015accurate} apply the Fourier phase-shifting theory to improve matching accuracy. Tao et al. \cite{tao2013depth} examine two depth cues derived from correspondence and defocus in LF data.

To address occlusion-related errors, Wang et al. \cite{wang2016depth} expand on \cite{tao2013depth} by developing a model that incorporates occlusion effects in LFs and propose an anti-occlusion algorithm that analyzes positional relationships between subviews and integrates occlusion knowledge into stereo matching. Sheng et al. \cite{sheng2018occlusion} devise a depth estimation algorithm that combines epipolar plane images (EPIs) with local depth consistency to mitigate occlusion. Zhang et al. \cite{2016SPO} consider the structure of EPIs and propose a local depth estimation method using a spinning parallelogram operator (SPO). By utilizing regions segmented by the SPO in EPIs, they maximize distribution distances to locate depth lines. Chen et al. \cite{chen2018accurate} propose a depth estimation framework that regularizes the label confidence map and edge strength weights by detecting partially occluded boundary regions (POBR) and applying shrinkage/reinforcement operations. Liu et al. \cite{liu2017iterative} introduce an iterative scheme for scene depth reconstruction based on a 4D LF, including a fidelity term for matching and penalty terms for gradient and classification components. Recently, Liu et al. \cite{LIU2023109042} map pixel differences between matching regions to Gaussian functions, reducing the values of pixels with larger differences in occluded and noisy areas, thereby achieving anti-occlusion resistance and noise reduction.

In contrast to these traditional optimization methods, the proposed method employs implicit neural representations for continuous disparity modeling. This method effectively utilizes the spatial-angular consistency of LF data, overcoming the limitations of discrete data.

\subsection{Deep learning-based methods}
Advances in deep learning have increased attention to LF depth estimation, which can be categorized into supervised and unsupervised methods. Supervised methods rely on labeled depth to learn mappings from LF inputs to depth values, focusing on network architecture design. In contrast, unsupervised methods infer depth directly from LF data by leveraging its intrinsic structure and designing task-specific loss functions.

In supervised methods, early research typically predicted disparity information from LF data by designing feature extractors and employing `end-to-end' network architectures. Heber et al. \cite{heber2016convolutional} introduced a CNN-based supervised method for depth estimation. Shin et al. \cite{shin2018epinet} extracted features from LF data in various directions (horizontal, vertical, and diagonal) and combined them using CNNs for depth estimation. Alperovich et al. \cite{alperovich2018light} developed a fully convolutional autoencoder for simultaneous depth and reflectance estimation. Li et al. \cite{2020EPI} introduced an Oriented Relation Module that captures spatial relationships in EPIs and enhances training effectiveness through refocusing-based data augmentation. Tsai et al. \cite{LFaTTNet} proposed a viewpoint-selection module that generates attention maps to assess the importance of each view, improving the utilization of LF data. Huang et al. \cite{Huang2021ICCV} developed a lightweight convolutional network that integrates cost volumes and attention modules, enabling effective inference of LF inputs at varying angles and enhancing recovery in occluded regions. Li et al. \cite{li2021lightweight} created a model for disparity estimation, employing physics-based multiscale cost-volume convolutional aggregation and an edge-guided subnetwork to enhance geometric detail recovery near edges and overall performance. Wang et al. \cite{wang2022occlusion} developed an occlusion-aware matching cost constructor that utilizes a series of convolutions with specific dilation rates, efficiently integrating pixels without shifting operations. At the same time, Wang et al. \cite{DistgDisp} proposed a generic mechanism to disentangle these coupled pieces of information for LF disparity estimation. Chao et al. \cite{chao2023learning} proposed a method for learning disparity distribution by constructing an interpolation-based cost volume at the sub-pixel level and designing an uncertainty-aware focal loss based on Jensen-Shannon divergence.

In unsupervised methods, Peng et al. \cite{peng2018unsupervised} introduce a CNN-based approach for explicit depth estimation, using a combined loss function to align the disparity between warped sub-aperture images and the central view. Later, Peng et al. \cite{peng2020zero} propose a zero-shot learning-based framework for LF depth estimation. Iwatsuki et al. \cite{iwatsuki2022unsupervised} redesign the loss function to indirectly measure disparity map accuracy by evaluating reprojection errors among the input LF views, and design pixel-wise weights and an edge loss to assess reprojection errors in the presence of occlusions. Zhou et al. \cite{8858033} utilize geometrically constrained EPIs to train networks for depth reconstruction without requiring ground-truth depth labels. Furthermore, Zhou et al. \cite{zhou2023beyond} introduce an unsupervised LF depth estimation framework with explicit occlusion detection based on EPI geometry. Additionally, Li et al. \cite{li2023opal} propose the Occlusion Pattern Aware Loss (OPAL), which successfully extracts and encodes the general occlusion patterns inherent in the LF for loss calculation.

While supervised learning relies on precise depth labels for training, unsupervised learning avoids the need for labels but requires complex loss functions and large datasets (except for zero-shot methods), along with extended training times. In contrast, the method proposed in this paper, based on implicit neural representation, operates without large-scale datasets or specific camera models, offering greater adaptability and efficiency.

\subsection{Neural fields}
The neural radiance field \cite{2020NeRF} represents the scene to be reconstructed as a continuous implicit function parameterized by a neural network, achieving high-quality visual quality and inspiring numerous subsequent works \cite{barron2021mip, barron2022mip, verbin2022ref, muller2022instant, chen2022tensorf,chen2023factor}. Among these, mip-NeRF \cite{barron2021mip} reduces the multiscale representation of image aliasing in the neural radiance field. Mip-NeRF 360 \cite{barron2022mip} extended mip-NeRF to address poor performance in unbounded scene representations. Ref-NeRF \cite{verbin2022ref} addresses specular and reflection issues in novel view synthesis. The instant-NGP \cite{muller2022instant} model enhances the training speed of neural radiance field models. TensorRF \cite{chen2022tensorf} decomposes the radiance field tensor into multiple lower-order tensor components to obtain a more compact scene representation. Factor Fields \cite{chen2023factor} is a unified framework based on coordinate-input networks, allowing the creation of powerful new signal representations and achieving improvements in approximation quality, compactness, and training time.

Chugunov et al. \cite{chugunov2022implicit} tackle the challenge of obtaining high-fidelity depth maps in everyday photography by leveraging small baseline parallax from hand tremors, along with low-resolution LiDAR data. In a subsequent study \cite{chugunov2023shakes}, they propose an unsupervised approach that estimates high-quality depth and camera motion directly from 12-megapixel RAW frames and gyroscope data, relying on natural hand tremors for parallax cues. Although both methods aim to acquire depth from small baseline multiviews, they face challenges when applied to LF data. The minimal baselines between viewpoints and the equidistant sampling of the viewpoint plane result in the disparity between any two views being a multiple of the unit disparity relative to the viewpoint baseline. This constraint weakens effective disparity information and complicates disparity estimation using projection models. In the domain of Neural Disparity Refinement, Fabio et al. \cite{tosi2024neural} propose a neural network-based method to refine noisy disparity maps generated by both traditional and deep learning stereo algorithms, aiming to improve depth estimation accuracy across various scenarios.

Our method's core idea involves establishing a differentiable forward process that allows the neural network to learn, through backpropagation, to map the coordinates of the central view to the corresponding disparities within the scene. Unlike supervised or unsupervised methods, our approach is optimization-based, as it relies on an optimization process to iteratively refine the disparity estimation without requiring a training dataset.

\section{Method}
To ensure generality, the proposed method is demonstrated using the central viewpoint depth and possesses the ability to reconstruct depth information from alternative viewpoints. Additionally, for LFs, there exists a one-to-one correspondence between the scene depth and disparity, which accurately represents the 3D spatial structure of the scene. 

\subsection{Neural disparity field representation}
In contrast to traditional implicit representations of 3D shapes, disparity refers to a two-dimensional (2D) geometric manifold in 3D space that is associated with the depth of the scene. It describes the differences between viewpoints instead of directly representing 3D geometric shapes. The implicit continuous parameterization of NDFs allows the modeling of continuous disparities, making it more appropriate for describing continuous variations in disparity within LFs compared to the conventional 2D matrix representations.

The disparity inherent in a scene within an LF is parameterized as a continuous 2D implicit function using neural fields, termed NDFs. This is achieved by approximating the continuous 2D implicit function using a multiresolution grid and the multilayer perceptrons (MLPs) network $\mathbf{F}_\mathbf{\Theta}$ with activation functions, defined as
\begin{equation}
\mathbf{F}_\mathbf{\Theta}: \mathbf{x} \rightarrow {d(\mathbf{x})},
\label{Eq1}
\end{equation}
where the network takes 2D coordinate positions $\mathbf{x}=(x, y)$ as input and outputs the disparity ${d(\mathbf{x})}$ at that point. By optimizing the network weights $\mathbf{\Theta}$, this representation maps each input 2D coordinate $\mathbf{x}$ to its corresponding disparity ${d(\mathbf{x})}$.

\subsection{Inverse problem solving for disparity reconstruction}
In the LF, disparity information is shared across various subviews, allowing viewpoint transitions through disparity propagation. In the two-plane four-parameter LF, assuming that occlusion and camera noise are disregarded, the relationship between image $L_{(u_0, v_0)}(\mathbf{x})$ at the central viewpoint $(u_0, v_0)$ and image $L_{(u_i, v_j)}(\mathbf{x})$ at viewpoint $(u_i, v_j)$ can be described as follows: 
\begin{equation} 
L_{(u_0, v_0)}\left(\mathbf{x}\right)=L_{(u_i, v_j)}\left(\mathbf{x}+\Delta \cdot d(\mathbf{x})\right), 
\label{Eq2} 
\end{equation} 
where $\Delta =\left(u_i-u_0, v_j-v_0\right)$ represents the distance between viewpoints $(u_i, v_j)$ and $(u_0, v_0)$ in the $u$ and $v$ directions.

Certain factors such as occlusion and camera noise render (\ref{Eq2}) invalid for real-world scenarios. Therefore, this study introduces the propagation mask $m_{(u, v)}(\mathbf{x})$, with $m_{(u, v)}(\mathbf{x})=0$ in cases of occlusion or noise, and $m_{(u, v)}(\mathbf{x})=1$ otherwise. Based on this, the relationship between the central view image $L_{(u_0, v_0)}\left(\mathbf{x}\right)$ and the view image $L_{(u_i, v_j)}\left(\mathbf{x}\right)$ in the real scenes is adjusted as follows: 
\begin{equation} 
\label{Eq3} 
\begin{split} 
L_{(u_0, v_0)}(\mathbf{x})= m_{(u_i, v_j)}(\mathbf{x}) \cdot L_{(u_i, v_j)}\left(\mathbf{x}+\Delta \cdot d(\mathbf{x})\right). 
\end{split} 
\end{equation}

\subsubsection{Forward problem modeling}
The LF is regularly sampled on the viewpoint plane, ensuring that the pixels in the central view image are visible in at least half of the viewpoints, resulting in a stronger coupling between the central view and other views. Therefore, this study employs the disparity propagation process from other views to the central view in the LF to model disparity reconstruction. This process effectively utilizes the constraints of regular sampling on the viewpoint plane, while preserving the characteristics of the LF. This provides a robust forward problem model for disparity reconstruction.

To account for occlusion and camera noise in real scenes with an LF, a forward problem model for disparity reconstruction under angular consistency was established as follows:
\begin{footnotesize}
\begin{equation} \label{Eq5}
L(d(\mathbf{x})) = \frac{\sum_{(u_i,  v_j)\in \phi} \left(m_{(u_i,  v_j)}(\mathbf{x})  \cdot L_{(u_i,  v_j)}\left(\mathbf{x}+\Delta \cdot d(\mathbf{x})\right)\right)}{\sum_{(u_i, v_j)\in \phi} m_{(u_i, v_j)}(\mathbf{x})},
\end{equation}
\end{footnotesize}\\
where $\phi$ denotes the set of viewpoints in the LF data, excluding the central viewpoint.

\subsubsection{Inverse problem solving}
\paragraph{Loss functions}
Traditional stereo-matching algorithms often utilize region-based matching windows instead of individual pixels to enhance the robustness of the matching results\cite{shi2023matching}. To exploit the local structural information of the LF fully, this study utilized a combination of L1 and Mean Structural SIMilarity (MSSIM) losses \cite{wang2004image}. Total variation (TV) \cite{rudin1992nonlinear} regularization is introduced as a loss term to exploit the sparsity of disparities in the gradient domain.

This study utilized the L1 + $\alpha \cdot\left( 1-\text{MSSIM}\right)$ metric to evaluate the disparity between viewpoints that inadequately convey information and the central view image. This metric, denoted as distance $E_{(u_i, v_j)}\left(\mathbf{x}\right)$, is defined as follows: 
\begin{footnotesize}
\begin{equation} \label{Eq7}
\begin{split}
& E_{(u_i,  v_j)}\left(\mathbf{x}\right)= \left|L_{(u_0, v_0)}\left(\mathbf{x}\right)-L_{(u_i, v_j)}\left(\mathbf{x}+\Delta \cdot \hat{d}_{\Theta}(\mathbf{x})\right) \right| \\ & + \alpha \cdot\left(1- \text{MSSIM}\left(L_{(u_0, v_0)}\left(\mathbf{x}\right), L_{(u_i, v_j)}\left(\mathbf{x}+\Delta \cdot \hat{d}_{\Theta}(\mathbf{x})\right) \right)\right).
\end{split}
\end{equation}
\end{footnotesize}

Given that occluded pixels can transmit effective information from at least half of the viewpoints, this study proposes a simplified strategy to calculate disparity propagation masks. Initially, it assumes that the mask values of all viewpoints are set to 1, implying that all viewpoints are capable of providing effective information. Subsequently, by sorting the ray distances $E_{(u_i, v_j)}(\mathbf{x})$ of the viewpoints in the set $\phi$, the method selects half of the viewpoints whose ray distance errors are smaller than those of the other half. These selected viewpoints are then used to perform the backward propagation of loss. The calculation method for the propagation mask is expressed as follows:
\begin{equation}\label{eq_our}
m_{(u_i,  v_j)}(\mathrm{x})=\left\{\begin{array}{l}
1,  E_{(u_i,  v_j)}(\mathrm{x}) \leq E_{\text{median}}(\mathrm{x}) \\
0,  \text{ others }
\end{array}\right. ,
\end{equation}
where $E_{\text{median}}(\mathrm{x})$ represents the median of the ray distances in the viewpoint set $\phi$.

In general, this loss function is expressed as:
\begin{equation} \label{Eq6}
\begin{split}
\mathcal{L}= & \left|L_{(u_0, v_0)}\left(\mathbf{x}\right)-L\left( \hat{d}_{\Theta}\left(\mathbf{x}\right)\right)\right| + \\ & \alpha \cdot \left(1-\text{MSSIM}\left(L_{(u_0, v_0)}\left(\mathbf{x}\right), L\left( \hat{d}_{\Theta}\left(\mathbf{x}\right)\right)\right)\right)+\\ & \beta \cdot \Psi\left({\hat{d}_{\Theta}\left(\mathbf{x}\right)}\right),
\end{split}
\end{equation}
where $\hat{d}_{\Theta}\left(\mathbf{x}\right)$ represents the disparity predicted at point $\mathbf{x}$ by the network and $L(\hat{d}_{\Theta}(\mathbf{x}))$ is computed using (\ref{Eq5}). $\Psi(\hat{d}_{\Theta}(\mathbf{x}))$ denotes the regularization term, where TV regularization is employed. $\alpha$ and $\beta$ are the two hyperparameters.

This method effectively avoids the complex computation of disparity propagation masks while reducing the interference from occlusion noise. This loss function improves the accuracy of depth estimation in occluded scenes by prioritizing the effectiveness of information transfer.

\begin{figure}[!t]
\centering
\includegraphics[width=3.5in]{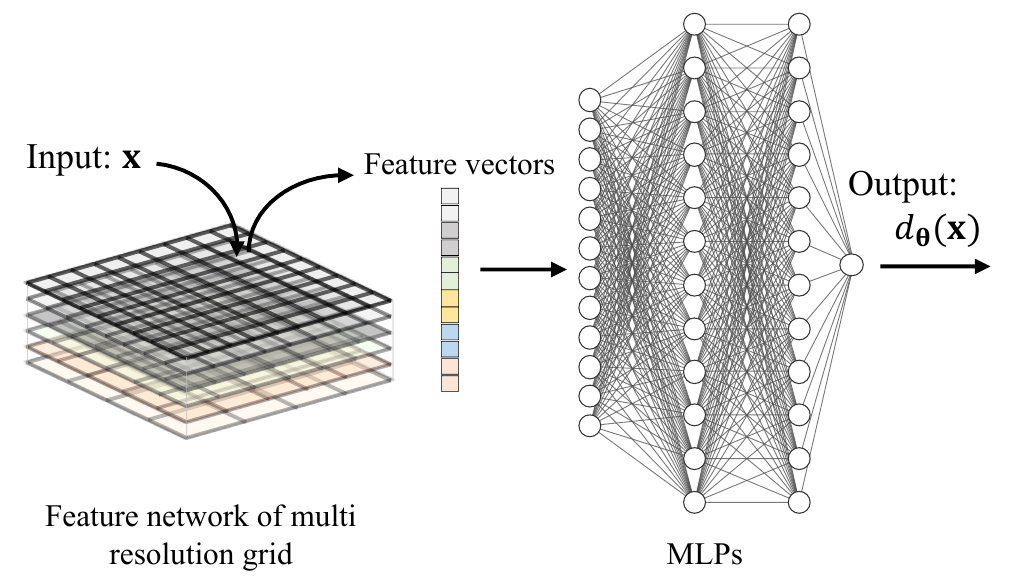}
\caption{NDF network architecture. The network consists of a multiresolution grid-feature module, which stores trainable Hash encoding feature vectors, and an MLP module composed of fully connected layers with LeakyReLU activation functions.}
\label{Fig_3}
\end{figure}

\paragraph{Network architecture}
Considering the disparity characteristics, adjacent pixels often have similar values, reflecting the smoothness of scene depth variations. This necessitates the precise capture of subtle changes using the NDF model. Additionally, the distribution of gradients in the disparity exhibits sparsity, with most pixels having gradients close to zero, which is significant only at the scene or object edges. This sparsity implies that the local pixels share the same features, obviating the need to learn unique feature descriptions for each independent pixel, thereby reducing the training time. Thus, this study constructs an NDF network based on the architecture provided in \cite{muller2022instant}. Its structure comprises a multiresolution grid feature module and an MLPs module, as depicted in Fig. \ref{Fig_3}. The multiresolution grid feature module, a trainable component, encompasses $L$ grids with resolutions linearly increasing from 32 to 128, and is utilized for feature vector storage. The MLPs module consisted of fully connected layers and LeakyReLU activation functions, comprising two hidden layers, each with 256 neurons.

To train the NDF, the following steps are executed:
\begin{enumerate}
    \item A set of 2D sampling points $\mathbf{x}$ is extracted from the coordinate system of the reference view.
    \item The multiresolution grid feature module is utilized to extract the corresponding features of the sampling points. These features are input into the MLPs module to predict the corresponding disparity $\hat{d}_{\Theta}\left(\mathbf{x}\right)$.
    \item The distance $E_{(u_i, v_j)}\left(\mathbf{x}\right)$ is calculated using (\ref{Eq7}) and is retained for those with $E_{(u_i, v_j)}\left(\mathbf{x}\right)$ smaller than the corresponding rays of the other half of the viewpoints.
    \item The loss of the retained rays is computed using (\ref{Eq6}). Subsequently, backward propagation is performed to endow the network with the ability to learn the disparity.
\end{enumerate}

\section{Experiments}
To comprehensively assess the performance of the proposed method, we conduct experiments on multiple LF datasets, including the HCI 4D LF dataset \cite{2013datasets}, the 4D LF Benchmark \cite{4DLFData}, and the UrbanLF-Syn dataset \cite{9810920}. Additionally, we test on the Stanford Lytro dataset \cite{2021StanfordLightField}, captured using a Lytro LF camera, and the (New) Stanford LF dataset \cite{Vaish2008}, which features a moving camera mounted on a gantry. All comparison methods are publicly available, and we compare the results with the state-of-the-art methods. Following the recommendations of the respective authors, the parameters of each algorithm are set to the optimal values for different datasets and scenes.

Experiments are conducted on an NVIDIA GeForce RTX 3080 with 10GB memory. The experimental settings include a grid feature module with six resolutions from 32 to 128, an MSSIM loss window size of 11 pixels, and hyperparameters $\alpha = 0.25$ and $\beta$ ranging from $1 \times 10^{-3}$ to $1 \times 10^1$ with slight variations. Gaussian noise with intensities ranging from $1 \times 10^{-2}$ to $1 \times 10^0$ is introduced during training. The training process is stopped after $8 \times 10^3$ iterations, processing LF data of size $9 \times 9 \times 512 \times 512$, which takes approximately 30 minutes on the RTX 3080. Using tiny-cuda-nn \cite{tiny-cuda-nn} reduces the runtime to less than 3 minutes.

\subsection{Algorithm analysis and evaluation using synthetic data experiments}
\subsubsection{Complex scene experiments}

\begin{figure}[!t]
\centering
\includegraphics[width=9cm]{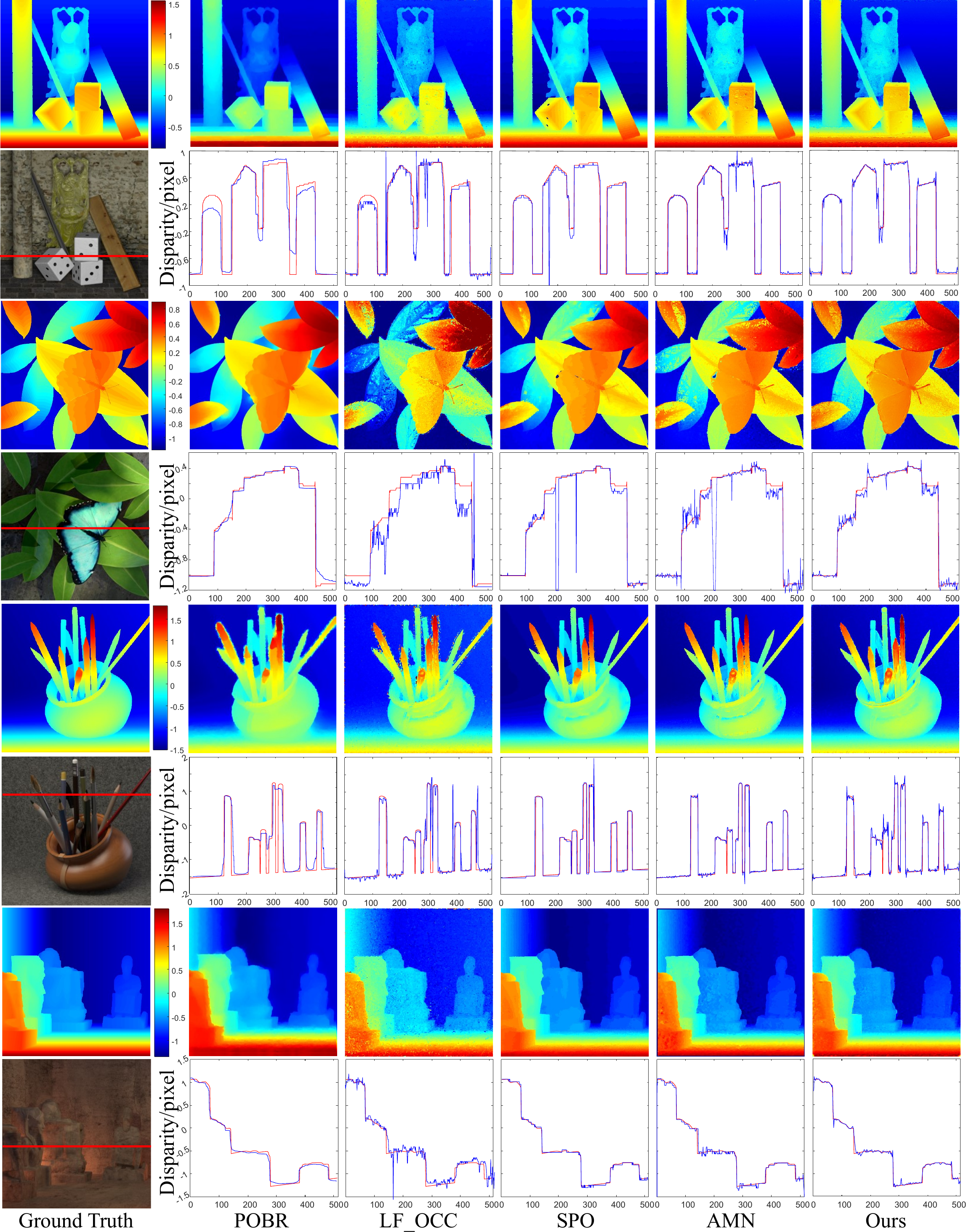}
\caption{Disparity results for complex synthetic scenes, with odd rows showing estimated disparity maps and even rows displaying profile curves (red line in reference view) comparing actual (red) and algorithm-estimated (blue) disparities. Scenes include from HCI \cite{2013datasets} (Buddha and Papillon) and 4D LF Benchmark \cite{4DLFData} (Pens and Tomb).}
\label{Fig_4}
\end{figure}

In this section, disparity estimation is performed using synthetic LF datasets \cite{2013datasets,4DLFData}, and compared with classical optimization-based methods that do not require a training dataset: POBR \cite{chen2018accurate}, which uses superpixel-based regularization; LF\_OCC \cite{wang2016depth}, a depth estimation method incorporating occlusion modeling; SPO \cite{2016SPO}, based on EPIs; and AMN \cite{LIU2023109042}, an adaptive matching norm method. The experimental results, including four scenes from the two datasets, are shown in Fig. \ref{Fig_4}.

\begin{figure}[!ht]
\centering
\includegraphics[width=8.5cm]{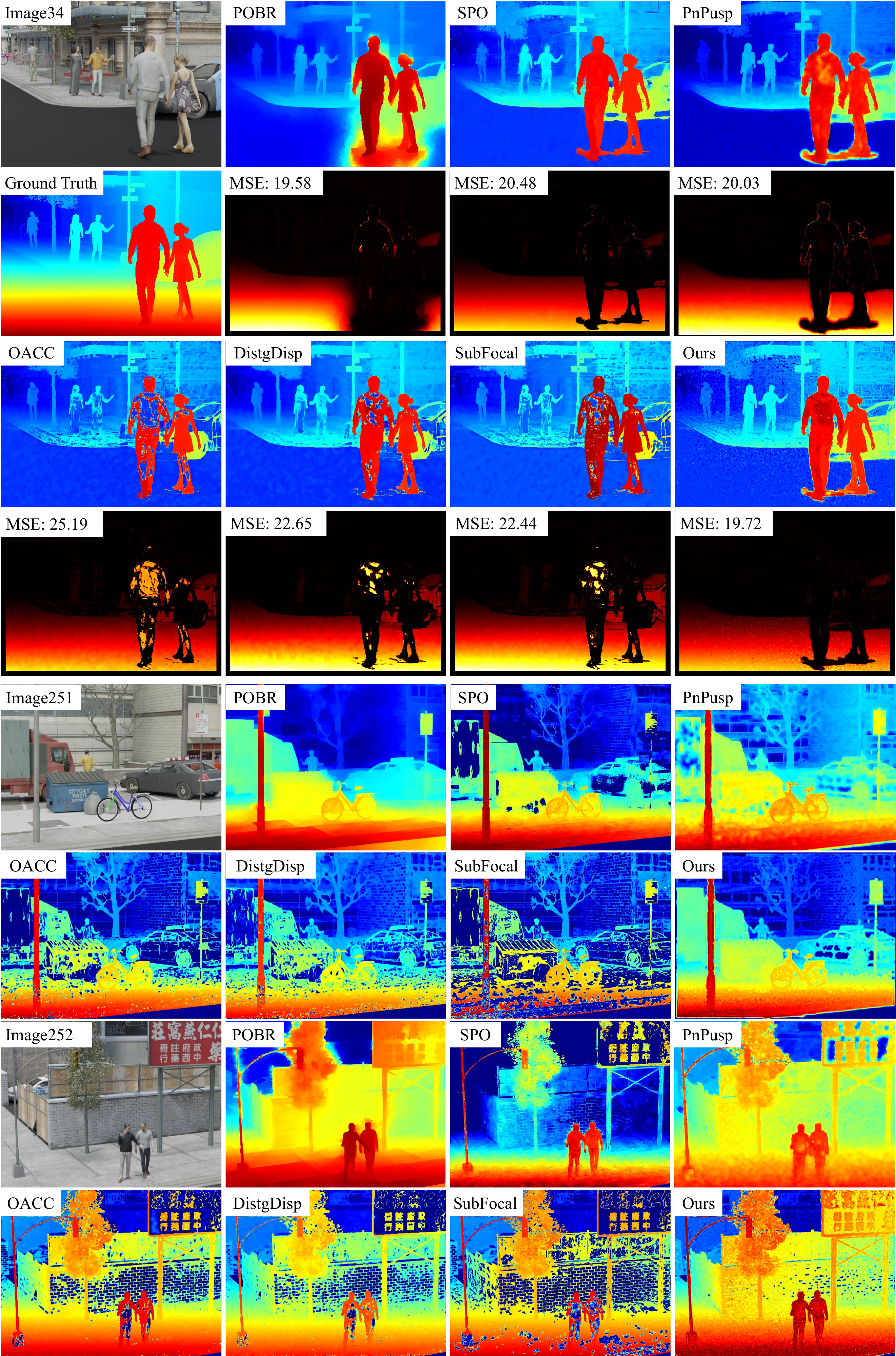}
\caption{Disparity results for complex synthetic scenes from the UrbanLF-Syn dataset \cite{9810920}. Image34 contains ground truth, and the corresponding MSE error maps are provided. Lower values are better. }
\label{Fig_5}
\end{figure}

In the Buddha scene, the prominent elements include a Buddha statue, slanted wooden planks, stone pillars, and dice. The intricate textures of Buddha statues and stone pillars pose challenges to the robustness of these algorithms. However, as shown in Fig. \ref{Fig_4}, all of the aforementioned algorithms effectively estimated the disparities of these elements. Slanted wooden planks tested the ability of the algorithms to handle inclined surfaces, where the proposed method outperformed POBR, producing smoother disparity predictions on slanted surfaces. Additionally, the black smooth regions on the dice present challenges to the algorithm performance, where the proposed method demonstrates notable advantages over SPO in handling textureless smooth areas. The profile graphs in the second row of Fig. \ref{Fig_4} show that the proposed method exhibited minimal oscillations, except for slight oscillations at the boundaries. In the Papillon scene, the primary elements are butterflies and leaves. The butterfly tail features a weakly textured black area and handling such a region is difficult. The proposed method successfully computed disparities in this area compared with SPO, and AMN. Furthermore, for the boundary areas of the leaves, LF\_OCC generates considerable noise owing to occlusion, whereas the proposed method exhibits advantages in handling such occluded regions. The Pens scene comprised multiple Pens arranged at various angles and depths. LF\_OCC introduces substantial noise at pen boundaries, whereas the proposed method and other comparative algorithms effectively reconstruct the disparities in this scene. The Tomb scene contains several stone statues, and the results reconstructed using the proposed method and SPO exhibit smoothness closer to the ground truth.

In addition, we compare our method with optimization-based methods, such as POBR and SPO, as well as publicly available supervised methods, including OACC \cite{wang2022occlusion}, DistgDisp \cite{DistgDisp}, and SubFocal \cite{chao2023learning}, and unsupervised methods like PnPusp, on the UrbanLF-Syn dataset. The experimental results are shown in Fig. \ref{Fig_5}. In this experiment, the selected scene is a virtual urban environment equipped with multiple light sources to simulate various lighting conditions. A camera array of 81 virtual cameras with identical configurations captures LF data. None of the algorithms in Fig. \ref{Fig_5} reconstruct the road in the scene, while our method recovers more details. In comparison, the disparity reconstruction from POBR loses detail and shows blocky artifacts in some regions. SPO is ineffective in certain smooth areas. Our method, with TV constraints, recovers more detail in these regions than SPO. Compared to supervised methods, these methods perform worse than expected on this dataset, while our method shows relative advantages.

Through comparative analysis, the proposed method demonstrates advantages in handling smooth and weakly textured areas.

\subsubsection{Specific scene experiments}
\begin{figure*}[!t]
\centering
\includegraphics[width=17cm]{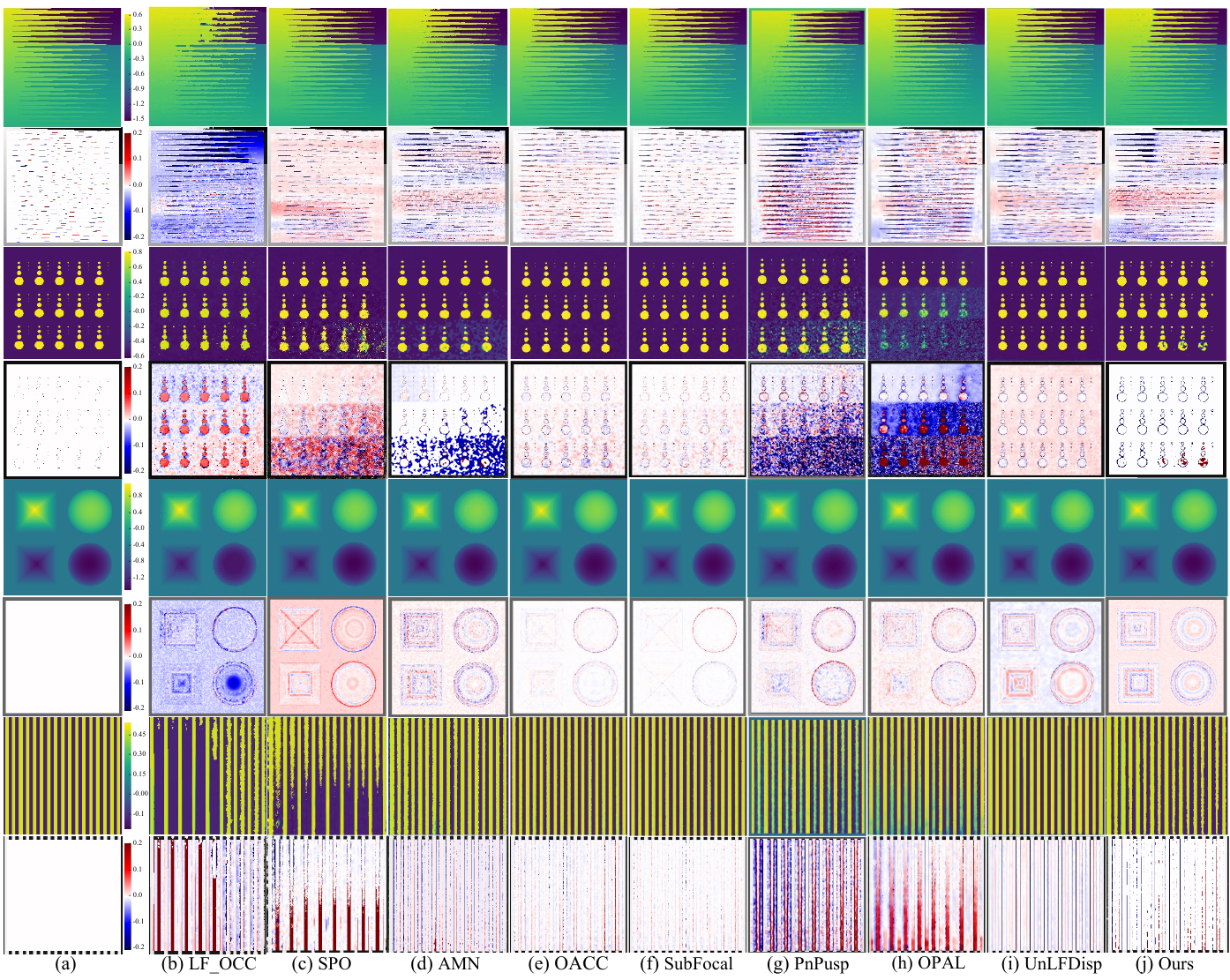}
\caption{Visual comparison of disparity estimation methods across different scenes. From top to bottom, rows correspond to Backgammon, Dots, Pyramids, and Stripes scene. For each scene, the top row shows the estimated disparity, and the bottom row shows the corresponding MSE. Results (b)-(j) are derived from original data in \cite{2017EvaluationDepth}.}
\label{Fig_6}
\end{figure*}

To gain a comprehensive understanding of the proposed method’s performance, we set up isolated challenges in specific scenarios: backgammons, dots, pyramids, and stripes. The Backgammon scene evaluated the algorithm's capability to handle fine structures and occlusion boundaries, while the Dots scene investigated the impact of varying levels of camera noise on the reconstruction of objects of different sizes. The Pyramids scene tested the algorithm's performance on convex, concave, circular, and planar geometrical structures, and the Stripes scene specifically assessed the influence of texture and contrast on occlusion boundary recognition. We conduct comparative experiments with state-of-the-art methods, where LF\_OCC, SPO, and AMN are optimization-based methods, OACC and SubFocal are supervised methods, and OPAL\cite{li2023opal}, PnPusp\cite{iwatsuki2022unsupervised}, and UnLFDisp\cite{zhou2023beyond} are unsupervised methods. The results are presented in Fig. \ref{Fig_6}.

\begin{table*}
\caption{Quantitative comparisons of results on synthetic datasets, presented with BP01(BadPix0.01), BP03(BadPix0.03), BP07(BadPix0.07), MSE ($\times$ 100), and Q25 metrics.}
\begin{center}
\renewcommand{\arraystretch}{1.25}
\setlength{\tabcolsep}{2.3mm}{
\begin{tabular}{c c|c|c|c|c|c|c|c|c|c}
\hline\hline
\multicolumn{2}{c|}{\multirow{2}*{Methods}} & \multicolumn{4}{c|}{Optimization-based} & \multicolumn{2}{c|}{Supervised}  & \multicolumn{3}{|c}{Unsupervised}\\
\cline{3-11}
& & LF\_OCC & SPO & AMN & Ours & OACC & SubFocal & PnPusp & OPAL & UnLFDisp\\
\hline 
\multirow{5}*{Backgammon} & BP01 & 88.839 & \textbf{49.940} & \textbf{43.599} & \textbf{54.878} & \textbf{21.613} & \textbf{12.473} & \textbf{53.840} & \textbf{49.773} & 55.137\\
\cline{2-11}
& BP03 & 39.826 & \textbf{8.639} & \textbf{13.354} & \textbf{16.378} & \textbf{6.640} & \textbf{4.281} & 27.753 & 17.268 & \textbf{12.188}  \\
\cline{2-11}
& BP07 & 19.006 & \textbf{3.782} & 6.913 & \textbf{6.656}& \textbf{3.931} & \textbf{3.194} &14.200& 8.832& \textbf{5.203} \\
\cline{2-11} 
& MSE & 21.587 & \textbf{4.587} & 12.287 & \textbf{8.929} & \textbf{3.938} & \textbf{3.667} &9.399& \textbf{5.177}& \textbf{6.857} \\
\cline{2-11} 
& Q25 & 1.680 & 0.572 & \textbf{0.367} & \textbf{0.539} & \textbf{0.190} & \textbf{0.090} &\textbf{0.440}&\textbf{0.448} & 0.567 \\
\hline 
\multirow{5}{*}{Dots} & BP01 & 65.762 & 58.079 & 21.658 & \textbf{4.538} & 21.022 & 15.511 &72.861& 87.631 & 54.987 \\
\cline{2-11}
& BP03 & 23.552 & 35.068 & 18.428 & \textbf{4.485} & \textbf{3.040} & \textbf{1.524} &48.081& 68.839 & \textbf{3.383}  \\
\cline{2-11}
& BP07 & 5.822 & 16.274 & 18.428 & \textbf{4.380} & \textbf{1.510} & \textbf{0.899}&29.864& 48.309 & \textbf{1.531}  \\
\cline{2-11}
& MSE & \textbf{3.301} & \textbf{5.238} & \textbf{2.936} & \textbf{5.433} & \textbf{1.418} & \textbf{1.301} &\textbf{4.769} &7.442 & \textbf{1.435}  \\
\cline{2-11} 
& Q25 & 0.577 & 0.973 & \textbf{0.003} & \textbf{0.003} & 0.198 & 0.134 &0.905& 2.100 & 0.762 \\
\hline 
\multirow{5}{*}{ Pyramids } & BP01 & 74.144 & 79.206 & \textbf{32.485} & \textbf{32.675} & \textbf{3.852} & \textbf{1.867} &\textbf{24.779}& \textbf{25.863} & \textbf{27.193} \\
\cline{2-11}
& BP03 & 20.495 & 6.263 & 5.643 & \textbf{2.853} & \textbf{0.536} & \textbf{0.411} &4.157& \textbf{2.616} & 3.425   \\
\cline{2-11}
& BP07  & 3.172 & 0.861 & 0.449 & \textbf{0.129} & 0.157 & 0.220 &0.659&0.430 & 0.269  \\
\cline{2-11}
& MSE & 0.098 & 0.043 & 0.023 & \textbf{0.016} & \textbf{0.004} & \textbf{0.005}& 0.021 &\textbf{0.016} & \textbf{0.016}  \\
\cline{2-11} 
& Q25 & 0.905 & 1.206 & \textbf{0.184} & \textbf{0.552} & \textbf{0.080} & \textbf{0.062} &\textbf{0.234}& \textbf{0.274}& \textbf{0.385}  \\
\hline 
\multirow{5}{*}{ Stripes } & BP01 & 53.669 & 21.879 & 13.499 & \textbf{8.389} & 15.244 & 9.386 &67.906& 54.561 & 22.341 \\
\cline{2-11}
& BP03 & 21.257 & 15.460 & \textbf{7.966} & \textbf{8.053}& \textbf{4.644} & \textbf{3.568} &43.372&25.512 & 8.513  \\
\cline{2-11}
& BP07 & 18.408 & 14.987 & \textbf{5.835} & \textbf{7.518} & \textbf{2.920} & \textbf{2.464} &24.708& 13.232& \textbf{5.123}  \\
\cline{2-11}
& MSE & 8.131 & 6.955 & \textbf{2.128} & \textbf{2.518} &  \textbf{0.845} & \textbf{0.821} &\textbf{2.371}& \textbf{1.362} & \textbf{1.798}  \\
\cline{2-11} 
& Q25 & 0.500 & 0.064 & 0.700 & \textbf{0.038} & 0.134 & 0.091 &0.706&0.487 & 0.209  \\
\hline 
\multirow{5}{*}{ Average } & BP01 & 70.604 & 52.276 & 27.810 & \textbf{25.12} & \textbf{15.433} & \textbf{9.809} &54.847& 54.457 & 39.9145  \\
\cline{2-11}
& BP03 & 26.283 & 16.358 & 11.348 & \textbf{7.942} & \textbf{3.715} & \textbf{2.446} &29.591& 28.559 & \textbf{6.877}\\
\cline{2-11}
& BP07 & 11.602 & 8.976 & 7.906 & \textbf{4.671} & \textbf{2.130} & \textbf{1.694} &17.358& 17.701 & \textbf{3.032} \\
\cline{2-11}
& MSE & 8.279 & \textbf{4.206} & 4.344 & \textbf{4.224} &  \textbf{1.551}  & \textbf{1.449} &\textbf{4.140}&\textbf{3.499} & \textbf{2.527}  \\
\cline{2-11} 
& Q25 & 0.916 & 0.704 & 0.313 & \textbf{0.283} &  \textbf{0.151}& \textbf{0.094} &0.571& 0.827& 0.481\\
\hline 
\multicolumn{2}{c|}{ Parameters (M) } & — & — & — & \textbf{0.159} & 5.01 & 5.06 & 5.12&1.047 & 59.1  \\
\hline\hline
\end{tabular}}
\label{tab1}
\end{center}
\end{table*}

Figures \ref{Fig_6} (b)-(j) correspond to the different algorithms. From the experimental results, it is evident that the proposed method successfully reconstructs disparities in various challenging scenarios. In the Backgammon scene, the proposed method performed remarkably well in handling complex textures, tilted surfaces, and textureless smooth regions, particularly when compared with LF\_OCC, and SPO, although it failed to estimate disparities at the upper-left crevice accurately. In the Dots scene, the proposed method reconstructs more foreground points in the upper-left corner compared to the lower-right corner, indicating a certain robustness to noise; however, as the noise increases, the proposed method loses its advantage. In Pyramids scene, all algorithms accurately reconstructed the convex and concave surfaces. The proposed method demonstrated some advantages in Stripes scene with lower demands for texture contrast. Table \ref{tab1} presents the quantitative analyses of the scenes depicted in Fig. \ref{Fig_6} based on BadPix (BP) and mean squared error (MSE) metrics, along with the sizes of the network model parameters. Scores that exceed those of our method are indicated in bold. The proposed method exhibits certain advantages over optimization-based methods and demonstrates competitiveness against unsupervised methods, although it does not perform as well as supervised methods. Regarding the network parameters metric, the proposed method is particularly efficient in lightweight design, with only 0.159M trainable parameters and lower memory usage, making it more suitable for resource-constrained environments than methods such as OACC, SubFocal, and UnLFDisp.

\subsection{Algorithm analysis and evaluation using real data experiments}
\begin{figure*}[!t]
\centering
\includegraphics[width=17cm]{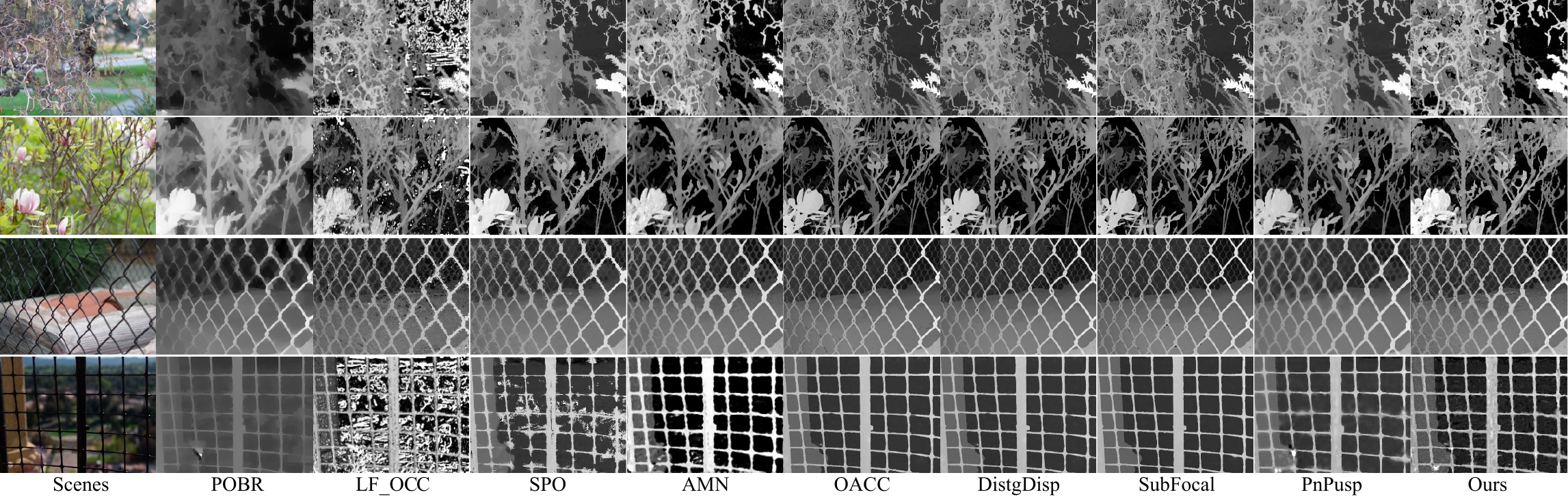}
\caption{Disparity estimation results for data from the Lytro Illum.}
\label{Fig_7}
\end{figure*}

\begin{figure}[!t]
\centering
\includegraphics[width=9cm]{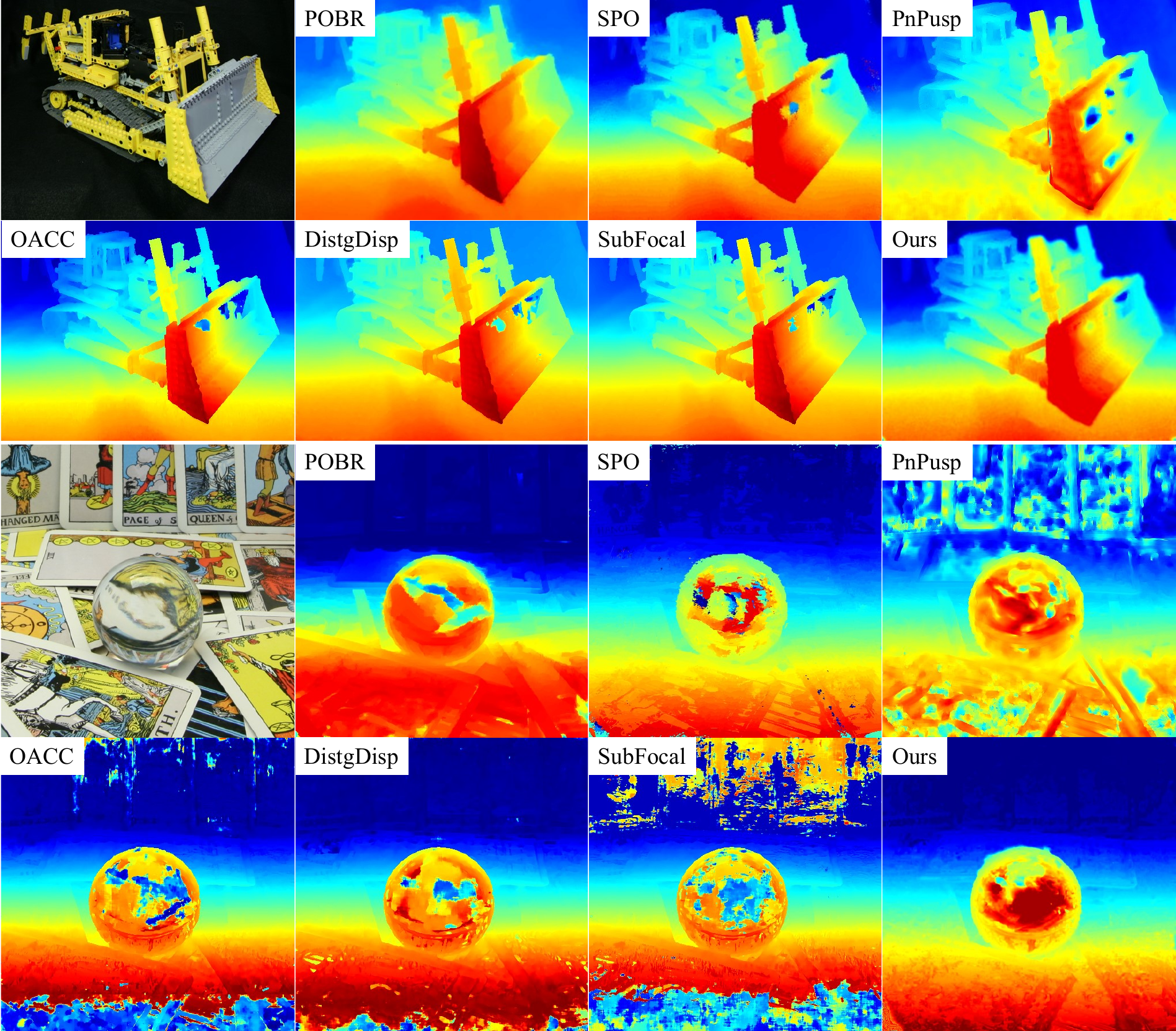}
\caption{Disparity estimation results for data from the moving camera. The first row shows a Lego Bulldozer, and the third row shows complex geometry: Tarot Cards and Crystal Ball (large angular extent).}
\label{Fig_8}
\end{figure}

Real LF data typically contain more noise, non-Lambertian radiance, and complex occlusion scenarios. We evaluate the performance of our method on real LFs captured by a moving camera \cite{Vaish2008} and the Stanford Lytro LF dataset \cite{2021StanfordLightField}. In this study, the proposed method is compared with POBR, LF\_OCC, SPO, AMN, OACC, DistgDisp, SubFocal, and PnPusp, as shown in Fig. \ref{Fig_7} and Fig. \ref{Fig_8}.

In Fig. \ref{Fig_7}, the first two rows show scenes with complex occlusion structures, suitable for evaluating the algorithm's performance in handling challenging scenarios. Both the proposed method and other methods accurately predict the intricate branches in the foreground. POBR exhibits over-smoothing, LF\_OCC shows significant errors in the background, SPO introduces multiple protrusion errors at the branch boundaries, and AMN loses detail due to excessive smoothing. The supervised learning methods perform the best. In the last two rows of Fig. \ref{Fig_7}, the foreground mainly consists of iron fences, which are accurately reconstructed by both the proposed method and other algorithms. However, in the background, the proposed method produces the least noise compared to POBR, LF\_OCC, SPO, AMN, and PnPusp, especially in the scene in the last row. In the third row of Fig. \ref{Fig_7}, the results from SubFocal show incorrect black spots within the gaps of the iron fence.

Fig. \ref{Fig_8} presents the disparity estimation results for LF data captured using a moving camera. In the Lego Bulldozer scene, which contains very complex geometry and textureless areas, most algorithms struggle significantly with the textureless regions. The proposed method, unlike POBR, estimates more disparity in these textureless areas. In the Tarot Cards and Crystal Ball scene, the cards serve as diffuse-textured objects at various depths and angles, and the scene includes large disparities with a textureless Crystal Ball. In this challenging setup, PnPusp, OACC, and SubFocal perform the worst. While none of the algorithms accurately estimate the disparity of the Crystal Ball, the proposed method achieves the best performance.

The experimental results demonstrate that the proposed method effectively predicts disparity structures in real scenes for LF data captured by both Lytro and moving cameras, reconstructing more details compared to other methods.

\subsection{Ablation studies}
\begin{figure}[!t]
\centering
\includegraphics[width=9cm]{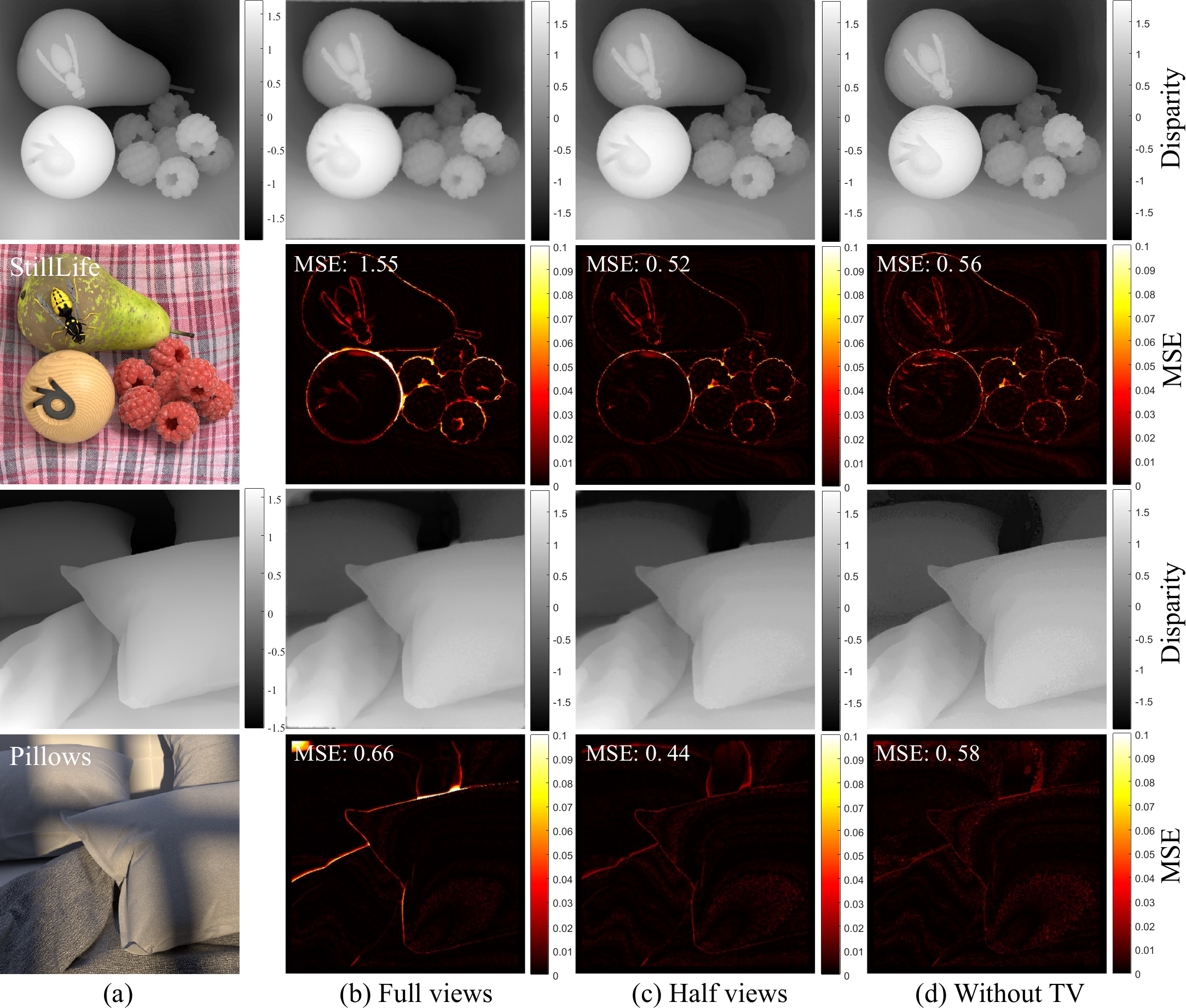}
\caption{The first row shows the disparity maps, while the second row (excluding (a)) presents the MSE maps. Column (b) uses all views, while column (c) uses half of the views as specified in \eqref{eq_our}. Column (d) uses half of the views and does not use TV}
\label{Fig_9}
\end{figure}

This section validates the effectiveness of the loss strategy in \eqref{eq_our}, where only half of the viewpoints are selected for backpropagation in each iteration. The effect of the TV constraint is also demonstrated. Fig. \ref{Fig_9} presents the results of the ablation study. In Fig. \ref{Fig_9}, columns (b) and (c) share the same parameters except for the number of viewpoints used. Specifically, column (b) uses all viewpoints, leading to significant errors at the boundaries. In contrast, column (c) uses only half of the viewpoints as specified by \eqref{eq_our}, significantly reducing boundary errors. This improvement is reflected in both the MSE maps and the performance metrics. Comparing columns (c) and (d), column (c) applies the TV constraint, while column (d) does not. The reconstruction results show that, in non-boundary texture regions of the scene, the disparity surfaces in column (c) are smoother, whereas column (d) exhibits more noise in these areas, as evident in the error maps.

\section{Discussion}
We conduct comparative experiments on multiple LF datasets to validate the performance of the proposed method in disparity estimation. The results demonstrate the effectiveness of the proposed method in addressing inclined surfaces, smooth textureless regions, and occlusions in intricate scenes. Specifically, when challenged with the weakly textured tail area of the butterfly and the occluded leaf boundaries in the Papillon scene, the proposed method achieves higher accuracy and smoother disparity estimates than BOPR, LF\_OCC, SPO, and AMN. Further validation is obtained through targeted experiments in scenes such as backgammon, dots, pyramids, and stripes, designed to test its ability to handle fine structures, noise, convex-concave geometries, and low-texture contrast scenarios. Although a slight decline in performance is observed under high-noise conditions, the proposed method achieves effective disparity reconstruction and preserves detail in most cases.

Additionally, experiments on the Stanford Lytro and moving camera dataset demonstrate the proposed method’s ability to discern disparity structures in noisy and complex occluded real-world scenes, excelling in background noise suppression and foreground edge reconstruction compared to other algorithms. While there remains a disparity gap compared to supervised or unsupervised methods relying on large datasets, the proposed method reconstructs greater disparity detail in challenging scenes.

Disparity describes the geometric structure of a scene, and disparity maps typically exhibit higher frequencies than natural images, especially along edges with disparity discontinuities. As a result, the proposed method exhibits oscillations in these regions. Future research should explore parameterized network models tailored for disparity estimation to improve performance. In summary, the proposed compact network model shows efficiency and robustness in handling weak textures, varying noise levels, and occlusion scenarios, offering a solution for precise LF depth estimation in practical applications.

\section{Conclusions}
We propose an iterative approach to reconstruct the NDF that utilizes neural network parameterization to achieve an implicit, continuous disparity function. The essential feature of this method is the implicit parameterization of the disparity by neural fields, which overcomes the resolution limitations inherent in traditional LF disparity estimation. Compared to conventional methods based on discrete disparity maps, the proposed approach reduces computational complexity while maintaining high precision and demonstrating superior robustness, particularly in handling complex scenes where occlusion and noise affect accurate disparity reconstruction. The proposed method employs a differentiable framework to construct a neural network model capable of modeling disparity continuously, thereby effectively avoiding the loss of detail caused by discrete representations. During the backpropagation optimization process, only a subset of rays is selected for propagation computation in the viewpoint direction, helping to mitigate occlusion effects and noise interference in the disparity estimation. Experimental validation demonstrates that the reconstructed NDF estimates disparity on both synthetic and real datasets.

\bibliography{ref}

\begin{thebibliography}{10}
\providecommand{\url}[1]{#1}
\csname url@samestyle\endcsname
\providecommand{\newblock}{\relax}
\providecommand{\bibinfo}[2]{#2}
\providecommand{\BIBentrySTDinterwordspacing}{\spaceskip=0pt\relax}
\providecommand{\BIBentryALTinterwordstretchfactor}{4}
\providecommand{\BIBentryALTinterwordspacing}{\spaceskip=\fontdimen2\font plus
\BIBentryALTinterwordstretchfactor\fontdimen3\font minus
  \fontdimen4\font\relax}
\providecommand{\BIBforeignlanguage}[2]{{%
\expandafter\ifx\csname l@#1\endcsname\relax
\typeout{** WARNING: IEEEtran.bst: No hyphenation pattern has been}%
\typeout{** loaded for the language `#1'. Using the pattern for}%
\typeout{** the default language instead.}%
\else
\language=\csname l@#1\endcsname
\fi
#2}}
\providecommand{\BIBdecl}{\relax}
\BIBdecl

\bibitem{2006LevoyLF}
M.~Levoy, ``Light fields and computational imaging,'' \emph{Computer}, vol.~39,
  no.~8, pp. 46--55, 2006.

\bibitem{2020NeRF}
B.~Mildenhall, P.~P. Srinivasan, M.~Tancik, J.~T. Barron, R.~Ramamoorthi, and
  R.~Ng, ``Nerf: Representing scenes as neural radiance fields for view
  synthesis,'' \emph{Communications of the ACM}, vol.~65, no.~1, pp. 99--106,
  2021.

\bibitem{NeRFPPF}
A.~Mittal, ``Neural radiance fields: Past, present, and future,'' \emph{arXiv
  preprint arXiv:2304.10050}, 2023.

\bibitem{NeFieldsinVisualComputingBeyond}
Y.~Xie, T.~Takikawa, S.~Saito, O.~Litany, S.~Yan, N.~Khan, F.~Tombari,
  J.~Tompkin, V.~Sitzmann, and S.~Sridhar, ``Neural fields in visual computing
  and beyond,'' in \emph{Computer Graphics Forum}.\hskip 1em plus 0.5em minus
  0.4em\relax Wiley Online Library, 2022, pp. 641--676.

\bibitem{levoy2023light}
M.~Levoy and P.~Hanrahan, ``Light field rendering,'' in \emph{Seminal Graphics
  Papers: Pushing the Boundaries, Volume 2}, 2023, pp. 441--452.

\bibitem{ng2005light}
R.~Ng, M.~Levoy, M.~Br{\'e}dif, G.~Duval, M.~Horowitz, and P.~Hanrahan, ``Light
  field photography with a hand-held plenoptic camera,'' Ph.D. dissertation,
  Stanford university, 2005.

\bibitem{xiao2013advances}
X.~Xiao, B.~Javidi, M.~Martinez-Corral, and A.~Stern, ``Advances in
  three-dimensional integral imaging: sensing, display, and applications,''
  \emph{Applied optics}, vol.~52, no.~4, pp. 546--560, 2013.

\bibitem{max1995optical}
N.~Max, ``Optical models for direct volume rendering,'' \emph{IEEE Transactions
  on Visualization and Computer Graphics}, vol.~1, no.~2, pp. 99--108, 1995.

\bibitem{wanner2013variational}
S.~Wanner and B.~Goldluecke, ``Variational light field analysis for disparity
  estimation and super-resolution,'' \emph{IEEE transactions on pattern
  analysis and machine intelligence}, vol.~36, no.~3, pp. 606--619, 2013.

\bibitem{2014PCA}
S.~Heber and T.~Pock, ``Shape from light field meets robust pca,'' in
  \emph{European Conference on Computer Vision}.\hskip 1em plus 0.5em minus
  0.4em\relax Springer, 2014, pp. 751--767.

\bibitem{jeon2015accurate}
H.-G. Jeon, J.~Park, G.~Choe, J.~Park, Y.~Bok, Y.-W. Tai, and I.~So~Kweon,
  ``Accurate depth map estimation from a lenslet light field camera,'' in
  \emph{Proceedings of the IEEE conference on computer vision and pattern
  recognition}, 2015, pp. 1547--1555.

\bibitem{tao2013depth}
M.~W. Tao, S.~Hadap, J.~Malik, and R.~Ramamoorthi, ``Depth from combining
  defocus and correspondence using light-field cameras,'' in \emph{Proceedings
  of the IEEE International Conference on Computer Vision}, 2013, pp. 673--680.

\bibitem{wang2016depth}
T.-C. Wang, A.~A. Efros, and R.~Ramamoorthi, ``Depth estimation with occlusion
  modeling using light-field cameras,'' \emph{IEEE transactions on pattern
  analysis and machine intelligence}, vol.~38, no.~11, pp. 2170--2181, 2016.

\bibitem{sheng2018occlusion}
H.~Sheng, P.~Zhao, S.~Zhang, J.~Zhang, and D.~Yang, ``Occlusion-aware depth
  estimation for light field using multi-orientation epis,'' \emph{Pattern
  Recognition}, vol.~74, pp. 587--599, 2018.

\bibitem{2016SPO}
S.~Zhang, H.~Sheng, C.~Li, J.~Zhang, and Z.~Xiong, ``Robust depth estimation
  for light field via spinning parallelogram operator,'' \emph{Computer Vision
  and Image Understanding}, vol. 145, pp. 148--159, 2016.

\bibitem{chen2018accurate}
J.~Chen, J.~Hou, Y.~Ni, and L.-P. Chau, ``Accurate light field depth estimation
  with superpixel regularization over partially occluded regions,'' \emph{IEEE
  Transactions on Image Processing}, vol.~27, no.~10, pp. 4889--4900, 2018.

\bibitem{liu2017iterative}
C.~Liu, J.~Qiu, and S.~Zhao, ``Iterative reconstruction of scene depth with
  fidelity based on light field data,'' \emph{Applied Optics}, vol.~56, no.~11,
  pp. 3185--3192, 2017.

\bibitem{LIU2023109042}
\BIBentryALTinterwordspacing
C.~Liu, L.~Shi, X.~Zhao, and J.~Qiu, ``Adaptive matching norm based disparity
  estimation from light field data,'' \emph{Signal Processing}, vol. 209, p.
  109042, 2023. [Online]. Available:
  \url{https://www.sciencedirect.com/science/article/pii/S0165168423001160}
\BIBentrySTDinterwordspacing

\bibitem{heber2016convolutional}
S.~Heber and T.~Pock, ``Convolutional networks for shape from light field,'' in
  \emph{Proceedings of the IEEE Conference on Computer Vision and Pattern
  Recognition}, 2016, pp. 3746--3754.

\bibitem{shin2018epinet}
C.~Shin, H.-G. Jeon, Y.~Yoon, I.~S. Kweon, and S.~J. Kim, ``Epinet: A
  fully-convolutional neural network using epipolar geometry for depth from
  light field images,'' in \emph{Proceedings of the IEEE conference on computer
  vision and pattern recognition}, 2018, pp. 4748--4757.

\bibitem{alperovich2018light}
A.~Alperovich, O.~Johannsen, M.~Strecke, and B.~Goldluecke, ``Light field
  intrinsics with a deep encoder-decoder network,'' in \emph{Proceedings of the
  IEEE conference on computer vision and pattern recognition}, 2018, pp.
  9145--9154.

\bibitem{2020EPI}
K.~Li, J.~Zhang, R.~Sun, X.~Zhang, and J.~Gao, ``Epi-based oriented relation
  networks for light field depth estimation,'' in \emph{British Machine Vision
  Conference (BMVC)}, 2020.

\bibitem{LFaTTNet}
Y.-J. Tsai, Y.-L. Liu, M.~Ouhyoung, and Y.-Y. Chuang, ``Attention-based view
  selection networks for light-field disparity estimation,'' in
  \emph{Proceedings of the 34th Conference on Artificial Intelligence (AAAI)},
  2020.

\bibitem{Huang2021ICCV}
Z.~Huang, X.~Hu, Z.~Xue, W.~Xu, and T.~Yue, ``Fast light-field disparity
  estimation with multi-disparity-scale cost aggregation,'' in
  \emph{Proceedings of the IEEE/CVF International Conference on Computer
  Vision}, 2021, pp. 6320--6329.

\bibitem{li2021lightweight}
Y.~Li, Q.~Wang, L.~Zhang, and G.~Lafruit, ``A lightweight depth estimation
  network for wide-baseline light fields,'' \emph{IEEE Transactions on Image
  Processing}, vol.~30, pp. 2288--2300, 2021.

\bibitem{wang2022occlusion}
Y.~Wang, L.~Wang, Z.~Liang, J.~Yang, W.~An, and Y.~Guo, ``Occlusion-aware cost
  constructor for light field depth estimation,'' in \emph{Proceedings of the
  IEEE/CVF conference on computer vision and pattern recognition}, 2022, pp.
  19\,809--19\,818.

\bibitem{DistgDisp}
Y.~Wang, L.~Wang, G.~Wu, J.~Yang, W.~An, J.~Yu, and Y.~Guo, ``Disentangling
  light fields for super-resolution and disparity estimation,'' \emph{IEEE
  Transactions on Pattern Analysis and Machine Intelligence}, 2022.

\bibitem{chao2023learning}
W.~Chao, X.~Wang, Y.~Wang, G.~Wang, and F.~Duan, ``Learning sub-pixel disparity
  distribution for light field depth estimation,'' \emph{IEEE Transactions on
  Computational Imaging}, vol.~9, pp. 1126--1138, 2023.

\bibitem{peng2018unsupervised}
J.~Peng, Z.~Xiong, D.~Liu, and X.~Chen, ``Unsupervised depth estimation from
  light field using a convolutional neural network,'' in \emph{2018
  International Conference on 3D Vision (3DV)}.\hskip 1em plus 0.5em minus
  0.4em\relax IEEE, 2018, pp. 295--303.

\bibitem{peng2020zero}
J.~Peng, Z.~Xiong, Y.~Wang, Y.~Zhang, and D.~Liu, ``Zero-shot depth estimation
  from light field using a convolutional neural network,'' \emph{IEEE
  Transactions on Computational Imaging}, vol.~6, pp. 682--696, 2020.

\bibitem{iwatsuki2022unsupervised}
T.~Iwatsuki, K.~Takahashi, and T.~Fujii, ``Unsupervised disparity estimation
  from light field using plug-and-play weighted warping loss,'' \emph{Signal
  Processing: Image Communication}, vol. 107, p. 116764, 2022.

\bibitem{8858033}
W.~Zhou, E.~Zhou, G.~Liu, L.~Lin, and A.~Lumsdaine, ``Unsupervised monocular
  depth estimation from light field image,'' \emph{IEEE Transactions on Image
  Processing}, vol.~29, pp. 1606--1617, 2020.

\bibitem{zhou2023beyond}
W.~Zhou, L.~Lin, Y.~Hong, Q.~Li, X.~Shen, and E.~E. Kuruoglu, ``Beyond
  photometric consistency: Geometry-based occlusion-aware unsupervised light
  field disparity estimation,'' \emph{IEEE Transactions on Neural Networks and
  Learning Systems}, 2023.

\bibitem{li2023opal}
P.~Li, J.~Zhao, J.~Wu, C.~Deng, Y.~Han, H.~Wang, and T.~Yu, ``Opal: Occlusion
  pattern aware loss for unsupervised light field disparity estimation,''
  \emph{IEEE Transactions on Pattern Analysis and Machine Intelligence}, 2023.

\bibitem{barron2021mip}
J.~T. Barron, B.~Mildenhall, M.~Tancik, P.~Hedman, R.~Martin-Brualla, and P.~P.
  Srinivasan, ``Mip-nerf: A multiscale representation for anti-aliasing neural
  radiance fields,'' in \emph{Proceedings of the IEEE/CVF International
  Conference on Computer Vision}, 2021, pp. 5855--5864.

\bibitem{barron2022mip}
J.~T. Barron, B.~Mildenhall, D.~Verbin, P.~P. Srinivasan, and P.~Hedman,
  ``Mip-nerf 360: Unbounded anti-aliased neural radiance fields,'' in
  \emph{Proceedings of the IEEE/CVF Conference on Computer Vision and Pattern
  Recognition}, 2022, pp. 5470--5479.

\bibitem{verbin2022ref}
D.~Verbin, P.~Hedman, B.~Mildenhall, T.~Zickler, J.~T. Barron, and P.~P.
  Srinivasan, ``Ref-nerf: Structured view-dependent appearance for neural
  radiance fields,'' in \emph{2022 IEEE/CVF Conference on Computer Vision and
  Pattern Recognition (CVPR)}.\hskip 1em plus 0.5em minus 0.4em\relax IEEE,
  2022, pp. 5481--5490.

\bibitem{muller2022instant}
T.~M{\"u}ller, A.~Evans, C.~Schied, and A.~Keller, ``Instant neural graphics
  primitives with a multiresolution hash encoding,'' \emph{ACM Transactions on
  Graphics (ToG)}, vol.~41, no.~4, pp. 1--15, 2022.

\bibitem{chen2022tensorf}
A.~Chen, Z.~Xu, A.~Geiger, J.~Yu, and H.~Su, ``Tensorf: Tensorial radiance
  fields,'' in \emph{European Conference on Computer Vision}.\hskip 1em plus
  0.5em minus 0.4em\relax Springer, 2022, pp. 333--350.

\bibitem{chen2023factor}
A.~Chen, Z.~Xu, X.~Wei, S.~Tang, H.~Su, and A.~Geiger, ``Factor fields: A
  unified framework for neural fields and beyond,'' \emph{arXiv preprint
  arXiv:2302.01226}, 2023.

\bibitem{chugunov2022implicit}
I.~Chugunov, Y.~Zhang, Z.~Xia, X.~Zhang, J.~Chen, and F.~Heide, ``The implicit
  values of a good hand shake: Handheld multi-frame neural depth refinement,''
  in \emph{Proceedings of the IEEE/CVF Conference on Computer Vision and
  Pattern Recognition}, 2022, pp. 2852--2862.

\bibitem{chugunov2023shakes}
I.~Chugunov, Y.~Zhang, and F.~Heide, ``Shakes on a plane: Unsupervised depth
  estimation from unstabilized photography,'' in \emph{Proceedings of the
  IEEE/CVF Conference on Computer Vision and Pattern Recognition}, 2023, pp.
  13\,240--13\,251.

\bibitem{tosi2024neural}
F.~Tosi, F.~Aleotti, P.~Z. Ramirez, M.~Poggi, S.~Salti, S.~Mattoccia, and
  L.~Di~Stefano, ``Neural disparity refinement,'' \emph{IEEE Transactions on
  Pattern Analysis and Machine Intelligence}, 2024.

\bibitem{shi2023matching}
L.~Shi, C.~Liu, D.~He, X.~Zhao, and J.~Qiu, ``Matching entropy based disparity
  estimation from light field data,'' \emph{Optics Express}, vol.~31, no.~4,
  pp. 6111--6131, 2023.

\bibitem{wang2004image}
Z.~Wang, A.~C. Bovik, H.~R. Sheikh, and E.~P. Simoncelli, ``Image quality
  assessment: from error visibility to structural similarity,'' \emph{IEEE
  transactions on image processing}, vol.~13, no.~4, pp. 600--612, 2004.

\bibitem{rudin1992nonlinear}
L.~I. Rudin, S.~Osher, and E.~Fatemi, ``Nonlinear total variation based noise
  removal algorithms,'' \emph{Physica D: nonlinear phenomena}, vol.~60, no.
  1-4, pp. 259--268, 1992.

\bibitem{2013datasets}
S.~Wanner, S.~Meister, and B.~Goldluecke, ``Datasets and benchmarks for densely
  sampled 4d light fields.'' in \emph{VMV}, vol.~13.\hskip 1em plus 0.5em minus
  0.4em\relax Citeseer, 2013, pp. 225--226.

\bibitem{4DLFData}
K.~Honauer, O.~Johannsen, D.~Kondermann, and B.~Goldluecke, ``A dataset and
  evaluation methodology for depth estimation on 4d light fields,'' in
  \emph{Asian conference on computer vision}.\hskip 1em plus 0.5em minus
  0.4em\relax Springer, 2016, pp. 19--34.

\bibitem{9810920}
H.~Sheng, R.~Cong, D.~Yang, R.~Chen, S.~Wang, and Z.~Cui, ``Urbanlf: A
  comprehensive light field dataset for semantic segmentation of urban
  scenes,'' \emph{IEEE Transactions on Circuits and Systems for Video
  Technology}, vol.~32, no.~11, pp. 7880--7893, 2022.

\bibitem{2021StanfordLightField}
S.~R. Abhilash, L.~Michael, S.~Raj, and W.~Gordon, ``Stanford lytro light field
  database,'' \url{http://lightfields.stanford.edu/LF2016.html}.

\bibitem{Vaish2008}
V.~Vaish and A.~Adams, ``The (new) stanford light field archive,''
  \url{http://lightfield.stanford.edu}, 2008, [Online].

\bibitem{tiny-cuda-nn}
\BIBentryALTinterwordspacing
T.~M\"uller, ``{tiny-cuda-nn},'' 4 2021. [Online]. Available:
  \url{https://github.com/NVlabs/tiny-cuda-nn}
\BIBentrySTDinterwordspacing

\bibitem{2017EvaluationDepth}
O.~{Johannsen}, K.~{Honauer}, B.~{Goldluecke}, A.~{Alperovich}, F.~{Battisti},
  Y.~{Bok}, M.~{Brizzi}, M.~{Carli}, G.~{Choe}, M.~{Diebold}, M.~{Gutsche},
  H.~{Jeon}, I.~S. {Kweon}, J.~{Park}, J.~{Park}, H.~{Schilling}, H.~{Sheng},
  L.~{Si}, M.~{Strecke}, A.~{Sulc}, Y.~{Tai}, Q.~{Wang}, T.~{Wang},
  S.~{Wanner}, Z.~{Xiong}, J.~{Yu}, S.~{Zhang}, and H.~{Zhu}, ``A taxonomy and
  evaluation of dense light field depth estimation algorithms,'' in \emph{2017
  IEEE Conference on Computer Vision and Pattern Recognition Workshops
  (CVPRW)}, 2017, pp. 1795--1812.

\end{thebibliography}
\bibliographystyle{IEEEtran}

\vfill

\end{document}